\documentclass[useAMS,usenatbib]{mn2e}
\usepackage{graphicx}
\usepackage{color}
\usepackage{soul}
\newcommand{\bec}[1]{\mbox{\boldmath $ #1$}}

\title[Dynamics of Wolf numbers based on nonlinear dynamo]{The dynamics of Wolf numbers based on nonlinear dynamos with magnetic helicity: comparisons with observations}

\author[Y. Kleeorin, N. Safiullin, N. Kleeorin, S. Porshnev, I. Rogachevskii, D. Sokoloff]
{Y. Kleeorin,$^{1}$
 N. Safiullin,$^{2}$
 N. Kleeorin,$^{3,4}$
 S. Porshnev,$^{2}$
 I. Rogachevskii,$^{3,4}$
 D.\, Sokoloff$^{5}$\\
  $^{1}$ Department of Physics,
        Ben-Gurion University of Negev, POB 653, 84105 Beer-Sheva, Israel\\
 $^{2}$ Department of Radio Electronic and Informational Technology,
        Ural Federal University, 19 Mira str., 620002 Ekaterinburg, Russia\\
 $^{3}$ Department of Mechanical Engineering,
        Ben-Gurion University of Negev, POB 653, 84105 Beer-Sheva, Israel\\
 $^{4}$ Nordita, KTH Royal Institute of Technology and Stockholm University,
        Roslagstullsbacken 23, SE-10691 Stockholm, Sweden\\
 $^{5}$ Department of Physics, Moscow
        State University, Moscow 119992, Russia}

\begin{document}



\maketitle


\begin{abstract}
We investigate the dynamics of solar activity using a nonlinear one-dimensional dynamo model and a phenomenological equation for the evolution of Wolf numbers. This system of equations is solved numerically. We take into account the algebraic and dynamic nonlinearities of the alpha effect. The dynamic nonlinearity is related to the evolution of a small-scale magnetic helicity, and it leads to a complicated behavior of solar activity. The evolution equation for the Wolf number is based on a mechanism of formation of magnetic spots as a result of the negative effective magnetic pressure instability (NEMPI). This phenomenon was predicted 25 years ago and has been investigated intensively in recent years through direct numerical simulations and mean-field simulations. The evolution equation for the Wolf number includes the production and decay of sunspots. Comparison between the results of numerical simulations and observational data of Wolf numbers shows a 70 \% correlation over all intervals of observation (about 270 years). We determine the dependence of the maximum value of the Wolf number versus the period of the cycle and the asymmetry of the solar cycles versus the amplitude of the cycle. These dependencies are in good agreement with observations.
\end{abstract}


\section{Introduction}

Solar activity has been studied from the time of Galileo who made first use of a telescope for the observation of sunspots. Since solar activity affects space weather, it is important to develop new methods for the prediction of solar activity. It is well established that solar dynamo plays an important role in explaning the quasi-periodic behaviour of the solar cycle
\citep{M78,P79,KR80,ZRS83,O03,RH04,BS05}. On the other hand, the solar dynamo alone cannot explain the formation of sunspots and active regions.

In spite of comprehensive observations of solar activity,
the mechanism for the formation of sunspots is still a subject of active discussions.
The traditional point of view is that the dynamo mechanism generates very strong, weakly non-uniform magnetic field at the bottom of the convective zone. Since this magnetic field is buoyant, it rises up, reaches the solar surface and creates a bipolar region
\citep{CSD95,DG06,CH08}. However, in recent years this idea has not been supported by helioseismology \citep{STR13,ZBK13}, by numerical simulations \citep{KKB14,PCM15,FM15} or by stability analysis \citep{ASR05}.

Another mechanism for the formation of sunspots is the kinematic effect of flux expulsion
\citep{C65,W66,TWB98,KKW10}. For instance, this can occur in large-scale convective circulations where the magnetic field is expelled from regions of fast motion. However, since this is a laminar effect, the role of a fully developed turbulence in this phenomenon is not clear.

An alternative mechanism for the formation of magnetic spots is associated with
the negative effective magnetic pressure instability (NEMPI) in strongly stratified turbulence.
This mechanism is based on the idea that a mean magnetic field causes a strong suppression of the total (kinetic and magnetic) turbulent
pressure. This phenomenon results in the effective magnetic pressure (the sum
of non-turbulent and turbulent contributions to the mean
magnetic pressure) becoming negative and a large-scale MHD
instability (i.e., NEMPI) can be excited.
This instability cannot produce any new magnetic flux. It only
redistributes the mean magnetic field in space so that the regions with super-equipartition magnetic fields become separated by regions with weak magnetic field. This
phenomenon has been investigated analytically
\citep{KRR89,KRR90,KMR93,KMR96,KR94,RK07} and detected in direct numerical simulations (DNS)
\citep{BKR10,BKR11,BKR12,BKR13,BGKR14,KBKR12,KBKR16,LKR13,LKR14,KBKMR13,JKR14}. NEMPI can also create bipolar regions in turbulence with an external coronal envelope \citep{WKR13,WKR16}. This mechanism is consistent with the idea that magnetic spots are formed in the upper part of the convective zone (see Brandenburg 2005). The bipolar regions are also formed in a two-layer turbulence with the dynamo generated field in the lower layer \citep{MKR14,JKR15}. The destruction of the bipolar regions is related to a magnetic reconnection \citep{JKR16}.

From an observational point of view, a key parameter characterising the solar activity
is the Wolf number, $W= 10 g +f$, where $g$ is the number of sunspot groups and $f$ is the total number of sunspots in the visible part of the sun. This parameter has been
measured over the span of three centuries \citep{G73,S89}. Based on ideas of NEMPI, we formulate a phenomenological budget equation for the evolution of Wolf numbers. This equation describes the competition between the rate of production and decay of sunspots. We take into account the following facts: (i) the rate of production and decay of sunspots depends on the mean magnetic field; (ii) the period of the dynamo waves is 11 years, while Wolf numbers change over much shorter times (up to 1-3 month). These conditions allow us to use a steady-state solution of the budget equation for the evolution of Wolf numbers together with numerical solutions of the dynamo equations. These dynamo equations take into account the dynamic and algebraic nonlinearities of the total alpha effect (the sum of kinetic and magnetic parts of alpha effect).
The dynamic nonlinearity is associated with the evolution equation for the current helicity
that is related to small-scale magnetic helicity
\citep{KR82,GD94,KRR95,KR99,KMR00,KMR02,KMR03,KMRS03,BF00,VC01,BB02,BS05}.

We compare the Wolf numbers obtained through numerical simulations with the observational data of Wolf numbers. We determine the governing parameters using an optimization approach to reach maximum correlation between numerical simulations and observations (up to 70 \%). We also note that there could be other mechanisms for dynamics of Wolf numbers related to purely statistical noise in the dynamo governing parameters caused by the averaging over a finite number of convective cells. Such an effect was considered by \cite{MSU08} and \cite{PSU12}.
Based on the approach developed in the present paper and using data assimilation techniques, it is possible to apply these methods for long-term predictions of solar activity (e.g., Wolf numbers). Predictions of solar activity have been discussed in a number of papers
\citep{DTG06,MO06,CCJ07,KA07,BT07,OS08,P08,U08,JD09,KK11,T09,T15}.
However, the problem of improving the predictions of solar activity is still a subject of active research.

\section{Dynamo model and Wolf number evolution}

We use the mean-field approach to study the solar dynamo. In particular,
the induction equation for the mean field reads:
\begin{equation}
\frac{\partial \bec{B}}{\partial t}= \bec{\nabla} {\bf \times}
(\bec{V} {\bf \times} \bec{B} + \bec{\cal E} - \eta \,
\bec{\nabla} {\bf \times} \bec{B}) ,
\label{E1}
\end{equation}
where ${\bec{ V}}$ is the mean velocity that describes the differential
rotation, $\eta$ is the magnetic diffusion due to the
electrical conductivity of the fluid, $\bec{\cal E} = \langle
{\bec{ u}} \times {\bec{ b}} \rangle$ is the mean electromotive
force, ${\bec{ u}}$ and $ {\bec{ b}}$ are fluctuations of the
velocity and magnetic field respectively, and angular brackets
denote averaging over an ensemble of fluctuations.
In isotropic turbulence, the electromotive force,
$\bec{\cal E} = \alpha(\bec{B}) {\bec B} - \eta_{_{T}} \bec{\nabla} {\bf
\times} \bec{B}$, includes the $\alpha$-effect and turbulent
magnetic diffusivity $\eta_{_{T}}$
\citep{M78,P79,KR80,ZRS83}.

\subsection{Dynamo equations}

We use spherical coordinates, $r, \theta, \phi$, and consider an
axisymmetric mean magnetic field, $ \bec{B} = B_\phi
\bec{e}_{\phi} + \bec{\nabla} {\bf \times} (A \bec{e}_{\phi})$.
We study the dynamo action in a thin convective shell.
To simplify the dynamo model, we
average the equations for $A$ and $B_\phi$ over the depth of the
convective shell, so that all quantities are functions of
colatitude $\theta$.
We neglect the curvature of the convective shell
and replace it by a flat slab. These assumptions allow us to obtain the
following non-dimensional dynamo equations:
\begin{eqnarray}
{\partial B_\phi \over \partial t} &=& G D \sin \theta {\partial
A \over \partial \theta} + {\partial ^2 B_\phi \over \partial
\theta^2} - \mu ^2 B_\phi ,
\label{eqB}\\
{\partial A \over \partial t} &=& \alpha B_\phi + {\partial^2 A
\over \partial \theta^2} - \mu^2 A ,
\label{eqA}
\end{eqnarray}
where the last terms, $-\mu^2 B_\phi$ and $-\mu^2
A$, in Eqs.~(\ref{eqB}) and~(\ref{eqA}) determine turbulent diffusive losses
in the radial direction.
The value of the parameter $\mu$ is determined by the following equation:
\begin{eqnarray}
\int_{2/3}^{1} {\partial^2 B_\phi \over \partial r^2} \,dr = - {\mu^2 B_\phi \over 3},
\label{M1}
\end{eqnarray}
where the radius $r$ that is measured in units of $R_\odot$, changes from $2/3$ to 1
inside the solar convective zone.
The value $\mu=3$ can mimic a convective
zone with a thickness of about 1/3 of the solar radius.
The differential rotation is characterized
by the parameter $G =\partial \Omega / \partial r =1$.
We use the standard profile of the kinetic part of the $\alpha$ effect
$\alpha(\theta) = \alpha_0 \sin^3 \theta \cos \theta$.
We are interested in dynamo waves propagating from the central solar
latitudes towards the equator \citep{P55}.
This implies that the dynamo number, $D$, is negative.

In Eqs.~(\ref{eqB}) and~(\ref{eqA}) we measure the length in
units of solar radius $R_\odot$, time in units of the turbulent
magnetic diffusion time $R_\odot^2 / \eta_{_{T}}$, the differential
rotation $\delta\Omega$ in units of the maximal value of $\Omega$, and
$\alpha $ is measured in units of the maximum value of the
kinetic part of the $ \alpha $-effect.
The toroidal magnetic field, $B_\phi$ is measured in the units of the equipartition field
$B_{\rm eq} = u \sqrt{4 \pi \rho}$, and the vector potential of the poloidal field $A$ is
measured in units of $R_{\alpha} R_\odot B_{\rm eq}$. The density $\rho$
is normalized by its value at the bottom of the convective zone, and
the integral scale of the turbulent motions $\ell$ and turbulent
velocity $u$ at the scale $\ell$ are measured in units of their
maximum values through the convective region. The
magnetic Reynolds number ${\rm Rm} = \ell u / \eta$
is defined using these maximal values, and the turbulent diffusivity
is $\eta_{_{T}} = \ell u / 3$.
The dynamo number is defined as $D = R_\alpha
R_\omega$, where $R_{\alpha} = \alpha_0 R_\odot / \eta_{_{T}}$ and
$R_\omega = (\delta \Omega) \, R_\odot^2 / \eta_{_{T}}$.

\subsection{Algebraic nonlinearity}

The total $\alpha$ effect is defined as the sum of the kinetic,
$\alpha^v= \chi^v \phi_{v}(B)$, and magnetic, $\alpha^m= \chi^c \phi_{m}(B)$, parts:
\begin{equation}
\alpha (r, \theta) = \chi^v \phi_{v}(B) + \chi^c \phi_{m}(B) .
\label{A3}
\end{equation}
The magnetic part of the $\alpha$ effect is related to the current helicity \citep{FPL75,PFL76}.
Here $\chi^v = - (\tau /3) \langle \bec{u}
\cdot(\bec{\nabla} {\bf \times} \bec{u}) \rangle$, while
$\chi^{c} = (\tau / 12 \pi \rho)
\langle \bec{b} \cdot (\bec{\nabla} {\bf \times} \bec{b})
\rangle$ and $\tau$ is the correlation time of the turbulent velocity field.
The magnetic helicity is related to the current helicity $\langle \bec{b} \cdot
(\bec{\nabla} {\bf \times} \bec{b}) \rangle$ in the
approximation of locally homogeneous turbulent convection \citep{KR99}.
The quenching functions $\phi_{v}(B)$ and $\phi_{m}(B)$ in
Eq.~(\ref{A3}) are determined by the following expressions:
\begin{eqnarray}
\phi_{v}(B) &=& (1/7) [4 \phi_{m}(B) + 3 L(B)] ,
\label{A4} \\
\phi_{m}(B) &=& {3 \over {8B^2}} [1 - \arctan (\sqrt{8} B) /
\sqrt{8} B],
\label{A5}
\end{eqnarray}
\citep{RK00,RK01}, where $ L(B) = 1 - 16
B^{2} + 128 B^{4} \ln (1 + 1/(8B^2))$.
These functions in the limiting cases are given by:
$\phi_{v}(B) = 1-(48/5)B^2$ and $\phi_{m}(B) = 1-(24/5)B^2$
for weak magnetic field, $B \ll 1/3$,
while $\phi_{v}(B) = 1/(4B^2)$ and $\phi_{m}(B) = 3/(8B^2)$ for
strong magnetic field, $B \gg 1/3$, where
$ \chi^v $ and $ \chi^c$ are measured in units of maximal
value of the $\alpha$-effect.
The function $\phi_{v}$ determines the algebraic quenching of the
kinetic part of the $\alpha $ effect.

The magnetic part $\alpha^m$ includes two kinds
of nonlinearities: the algebraic quenching, determined by the function
$\phi_{m}(B)$
\citep{FBC99,RK00,RK01} and the dynamic nonlinearity, which is determined by
Eq. (\ref{A6A}) discussed in the next section.
The algebraic quenching of the $\alpha$-effect
is caused by the effects of the mean magnetic field
on the electromotive force \citep{RK00,RK01,RK04}.

We average Eq.~(\ref{A3}) over the depth of the
convective zone, so that the first term in the averaged equation is
determined by the values taken at the middle part of
the convective zone, while for the second term we introduce a
phenomenological parameter $\sigma$:
\begin{equation}
\alpha (\theta) = \chi^v \phi_{v}(B) + \sigma \chi^c \phi_{m}(B) ,
\label{E31}\\
\sigma = \rho_\ast \int {\, dr \over \rho(r)} ,
\label{EE31}
\end{equation}
where the helicities and quenching functions are associated with a middle part of the convective zone. We also consider $\sigma > 1$ as a free parameter.

\subsection{Dynamical nonlinearity}

The function $\chi^c(\bec{B})$ is determined by a dynamical
equation that is derived from the conservation law for magnetic
helicity:
\begin{equation}
{\partial \chi^{c} \over \partial t} + \bec{\nabla} \cdot \bec{\Phi} + {\chi^c \over T} = -{1
\over 9 \pi \, \eta_{_{T}} \, \rho_\ast} \, (\bec{\cal E} {\bf \cdot}
\bec{B}) ,
\label{A6A}
\end{equation}
where $ \bec{\Phi} = -\kappa_{_{T}} \bec{\nabla} \chi^c$
is the turbulent diffusion flux of the magnetic helicity
\citep{KR99,KMR00,KMR02,KMR03,KMRS03,BS05}, and $T = \ell^2 / \eta$ is the relaxation time of
magnetic helicity.
When the large-scale magnetic helicity increases with magnetic field,
the evolution of the small-scale helicity should compensate for this growth,
because the total magnetic helicity is conserved.
The compensation mechanisms include the dissipation and transport of the magnetic helicity.
We rewrite the dynamical equation~(\ref{A6A}) for the function
$\chi^c(\bec{B})$ in non-dimensional form as
\begin{eqnarray}
&& {\partial \chi^c \over \partial t} + \left(T^{-1} + \kappa_{_{T}}
\mu^2\right)\chi^c = 2\left({\partial
A \over \partial \theta} {\partial B_\phi \over \partial \theta} + \mu^2 A B_\phi\right)
\nonumber\\
&&\quad - (\alpha/\xi) B^2 - {\partial \over \partial \theta} \left(B_\phi {\partial A \over \partial \theta} - \kappa_{_{T}} {\partial \chi^c \over \partial \theta} \right) ,
\label{E9}
\end{eqnarray}
where
\begin{eqnarray}
B^2 = \xi \left\{B_\phi^2 + R_\alpha^2 \left[\mu^2 A^2 + \left( {\partial A \over \partial \theta}\right)^2\right] \right\},
\label{X51}
\end{eqnarray}
and $\xi=2 (\ell/R_\odot)^2$.
Here we average Eq.~(\ref{A6A}) over the depth
of the convective zone, so that the average value of $T^{-1}$ is
\begin{eqnarray}
T^{-1} = H^{-1} \int T^{-1}(r) \,d r \sim {\Lambda_\ell \, R_\odot^2 \,
\eta \over H \, \ell^2 \, \eta_{_{T}}} .
\label{TE31}
\end{eqnarray}
Here $H$ is the depth of the convective zone,
$\Lambda_\ell$ is the characteristic scale of variations $\ell$,
$T(r) = (\eta_{_{T}} / R_\odot^{2}) (\ell^2 / \eta)$ is the non-dimensional
relaxation time of the magnetic helicity, and the quantities
$\Lambda_\ell , \, \eta, \, \ell$ in Eq.~(\ref{TE31}) are associated
with the upper part of the convective zone.

We should stress that the current helicity is an observed quantity. In particular,
more than twenty five years of continuous observations by several observational groups,
have resulted \citep{ZSP10} in a reconstruction of the current
helicity time-latitude (butterfly) diagram for more than one full solar
magnetic cycle (1988-2015). From this butterfly diagram it is
apparent that the current helicity affects the solar activity
cycle and follows a polarity law comparable with the Hale polarity
law for sunspots, although with more complicated behaviour
\citep{SBKR06,ZMKR06,ZMKR12}.

\subsection{Equation for dynamics of surface density of Wolf number}

The main observational quantity of the solar activity is the Wolf number. Based on the idea of NEMPI, we obtain a phenomenological budget equation for evolution of the surface density
of the Wolf number $\tilde W(t,\theta)$:
\begin{eqnarray}
{\partial \tilde W \over \partial t} = I(t,\theta) - {\tilde W \over \tau_s(B)} .
\label{B1}
\end{eqnarray}
This budget equation includes the rate of production of Wolf number, $\tilde W(t,\theta)$,
caused by the formation of sunspots,
\begin{eqnarray}
I(t,\theta) = {|\gamma_{\rm inst}| |B-B_{\rm cr}| \over \Phi_s} \Theta(B-B_{\rm cr}) ,
\label{B2}
\end{eqnarray}
and the rate of decay of sunspots. In particular,  the decay of sunspots that occurs in the nonlinear stage of the instability is described by the relaxation term, $-\tilde W /\tau_s(B)$. The functional form of the source function $I(t,\theta)$ determined by Eq.~(\ref{B2}) has been chosen by the following physical reasoning. It is assumed that the sunspots are produced by an instability (NEMPI) that has a threshold, i.e., the instability is excited only when $B>B_{\rm cr}$. This implies that the source $I(t,\theta)$ is proportional to a $\Theta$ function, namely $\Theta(B-B_{\rm cr})$, where the function $\Theta(x) = 1$, when $x>0$, and $\Theta(x) = 0$, when $x\leq 0$. The function $I(t,\theta)$ is the Wolf number change rate, and the characteristic time of change of the Wolf number is assumed to be identified with the characteristic time for excitation of the instability, $\gamma_{\rm inst}^{-1}$.
When $\gamma_{\rm inst}<0$, the rate of production, $I(t,\theta)$, vanishes.
This is the reason why $I(t,\theta) \propto \gamma_{\rm inst}$.
NEMPI does not produce any new magnetic flux. It redistributes the magnetic flux that is produced by the mean-field dynamo. The production term, $I(t,\theta)$, is also proportional to the maximum number of sunspots per unit area, that can be estimated as $\sim |B-B_{\rm cr}| /\Phi_s$, where $|B-B_{\rm cr}|$ is the magnetic flux per unit area that contributes to the sunspot formation and $\Phi_s$ is the magnetic flux inside a magnetic spot.

\begin{figure}
\centering
\includegraphics[width=9cm]{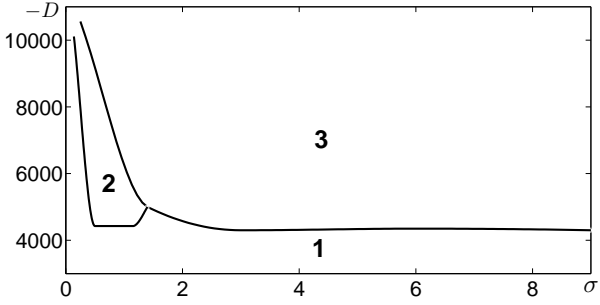}
\caption{\label{Fig1} The parameter range of the dynamo model: the dynamo number $-D$
versus the density parameter $\sigma$.}
\end{figure}

Since the decay time of sunspots $\tau_s(B)$ is about several weeks (up to 2-3 month), while the period of the dynamo wave is about 11 years, we can consider the steady-state solution of Eq.~(\ref{B1}): $\tilde W = \tau_s(B) \,I(t,\theta)$.
The Wolf number is determined as a surface integral:
\begin{eqnarray}
W &=& R_\odot^2 \, \int \tilde W(t,\theta) \sin \theta \, d\theta \, d\phi
\nonumber\\
&=& 2 \pi \,  R_\odot^2 \, \int \tau_s(B) \,I(t,\theta) \sin \theta \,d\theta .
\label{B3}
\end{eqnarray}
To choose the function $\tau_s(B)$ we take into account the fact that when the solar activity increases (decreases) the life time of sunspots increases (decreases). In particular, we choose $\tau_s(B)$ as
\begin{eqnarray}
\tau_s(B)=\tau_\ast \exp \left(C_s \, \partial B/\partial t\right) ,
\label{B4}
\end{eqnarray}
with $C_s= 1.8 \times 10^{-3}$ and $\tau_\ast \, \gamma_{\rm inst} \sim 10$, where the non-dimensional rate of the mean magnetic field, $\partial B/\partial t$, is measured in the units $\xi B_{\rm eq} / t_{\rm td}$, and $t_{\rm td}$ is the turbulent diffusion time.
Equation~(\ref{B4}) is mathematically based. We used other forms of the function $\tau_s(B)$, but the final results are weakly dependent on the form of this function.

We use estimates of governing parameters taken from models for the solar convective
zone, e.g., \cite{S74} or \cite{BT66}. In
the upper part of the convective zone, at depth $H_\ast=2 \times 10^7$ cm
(measured from the top), the parameters are
as follows: the magnetic Reynolds number ${\rm Rm} = 10^5$;
the integral scale of turbulence $\ell = 2.6 \times 10^7$ cm;
the characteristic turbulent velocity
in the integral scale of turbulence $u = 9.4 \times 10^4 $ cm s$^{-1}$;
the plasma density $\rho = 4.5 \times 10^{-7}$ g
cm$^{-3}$; the turbulent diffusion $\eta_{_{T}} = 0.8 \times 10^{12}$ cm$^2$ s$^{-1}$;
the equipartition mean magnetic field is $B_{\rm eq} = 220 $ G and
the non-dimensional relaxation time of the magnetic helicity is $T = 5 \times 10^{-3}$.
At depth $H_\ast= 10^9$ cm, the values of these parameters are
${\rm Rm}= 3 \times 10^7$; $\ell= 2.8 \times 10^8$ cm; $u= 10^4$ cm
s$^{-1}$; $\rho= 5 \times 10^{-4}$ g cm$^{-3}$; $\eta_{_{T}}= 0.9 \times 10^{12}$
cm$^2$ s$^{-1}$; $B_{\rm eq} = 800$ G and $T = 150$.
At the bottom of the convective
zone, at depth $H_\ast= 2 \times 10^{10}$ cm; ${\rm
Rm} = 2 \times 10^9$; $\ell = 8 \times 10^9$ cm;
$u = 2 \times 10^3 $ cm s$^{-1}$; $\rho = 2 \times 10^{-1}$ g
cm$^{-3}$; $\eta_{_{T}} = 5.3 \times 10^{12}$
cm$^2$s$^{-1}$; $B_{\rm eq} = 3000 $ G and $T = 10^7$.

\section{Results}

\begin{figure}
\centering
\includegraphics[width=9cm]{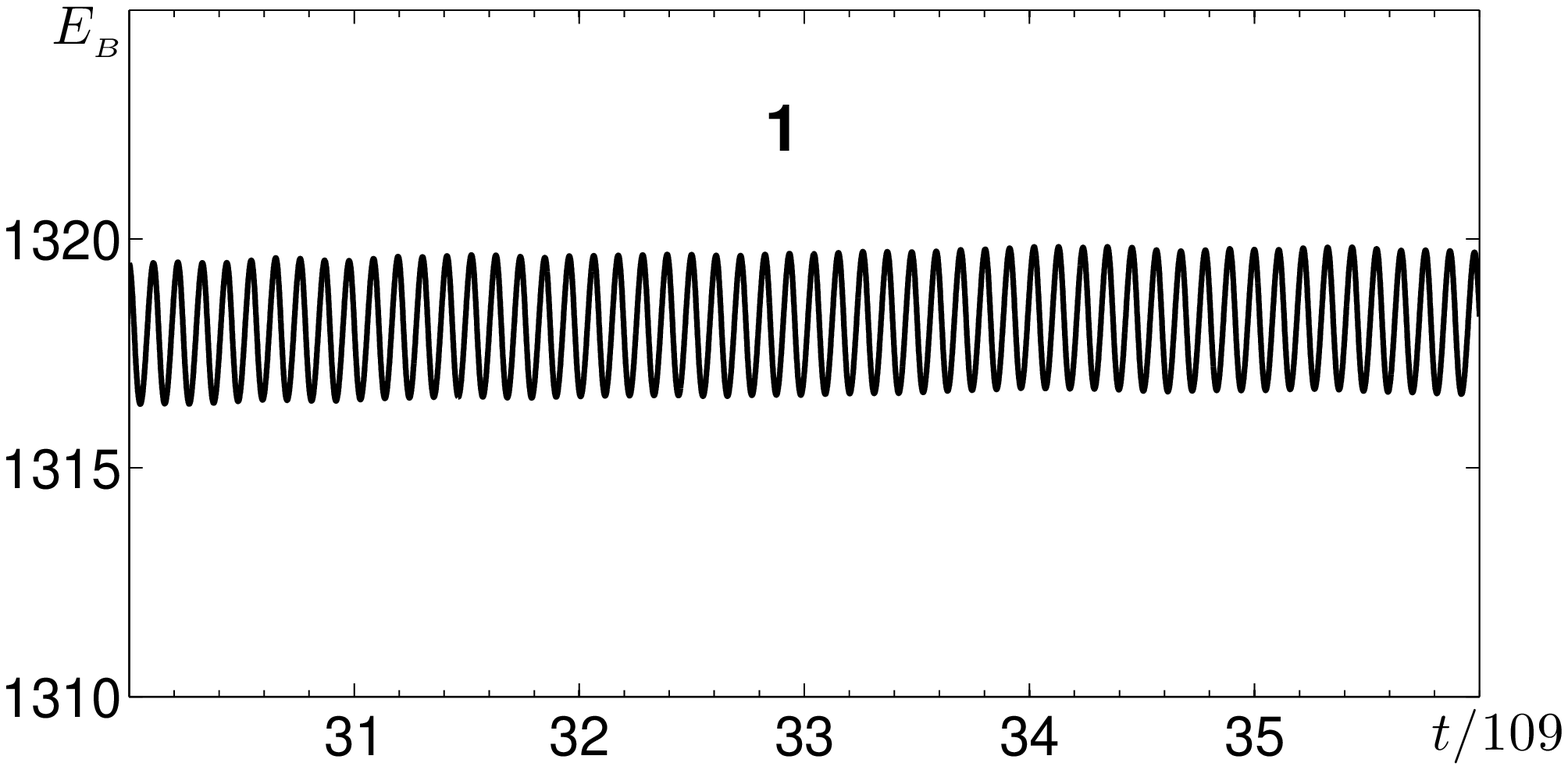}
\includegraphics[width=9cm]{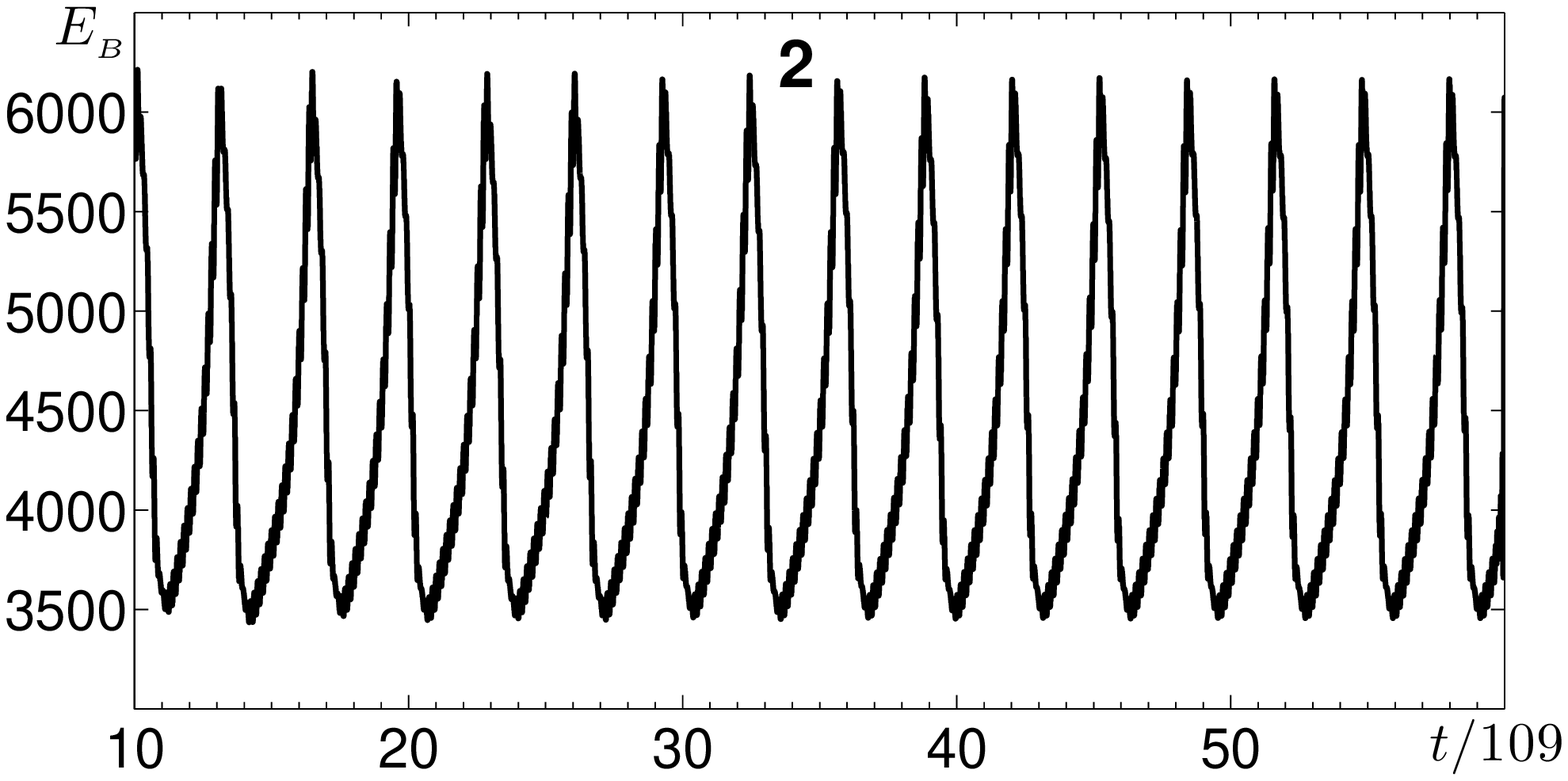}
\includegraphics[width=9cm]{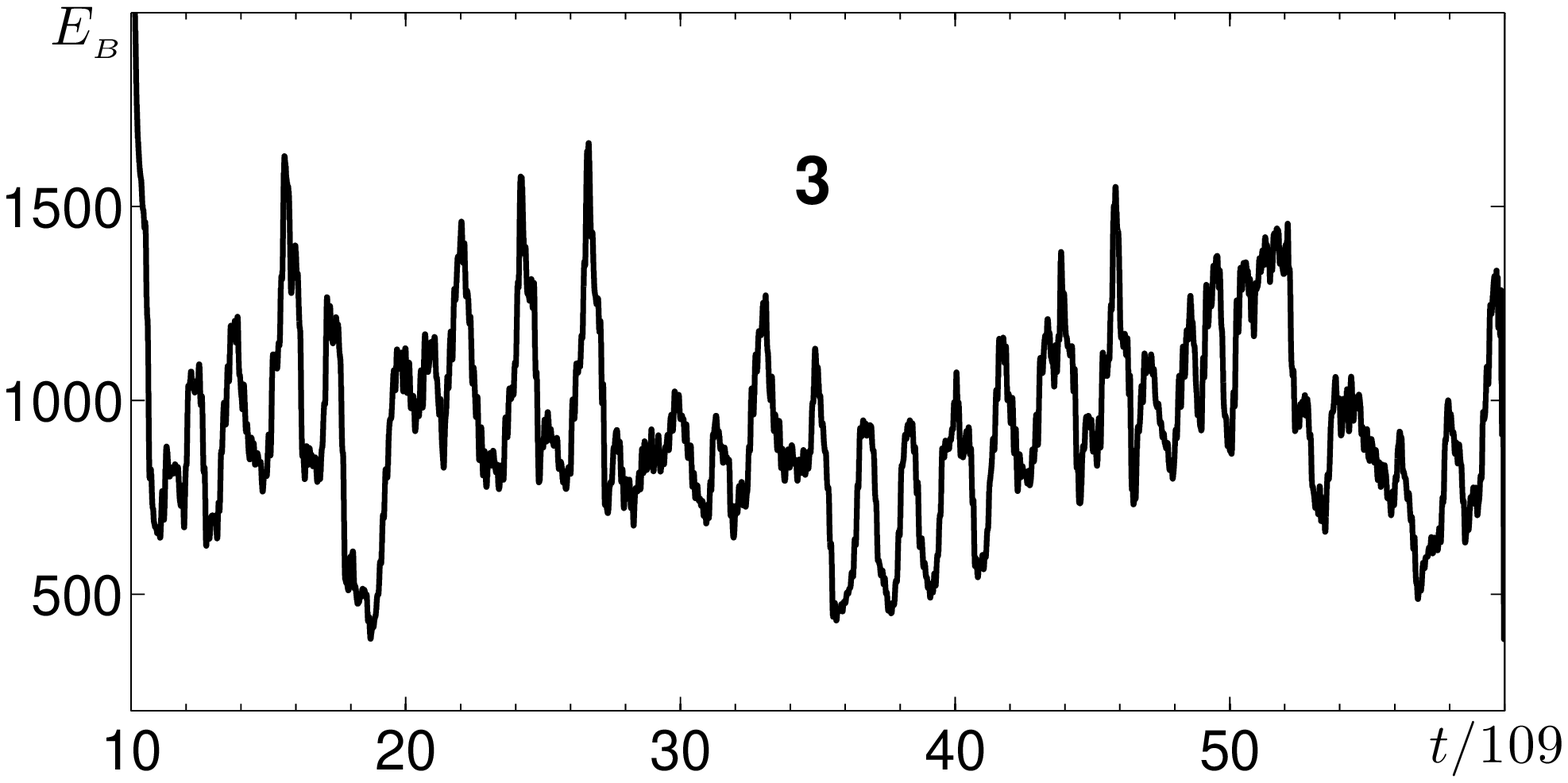}
\caption{\label{Fig2}  Three ranges of different behaviour of the nonlinear oscillations of magnetic energy:
1. The oscillations with a constant frequency (Fig.~2a, upper panel);
2. The oscillations with a low-frequency-modulation of amplitude and frequencies (Fig.~2b);
3. The chaotic behavior of magnetic energy (Fig.~2c, lower panel).}
\end{figure}

\begin{figure}
\centering
\includegraphics[width=9cm]{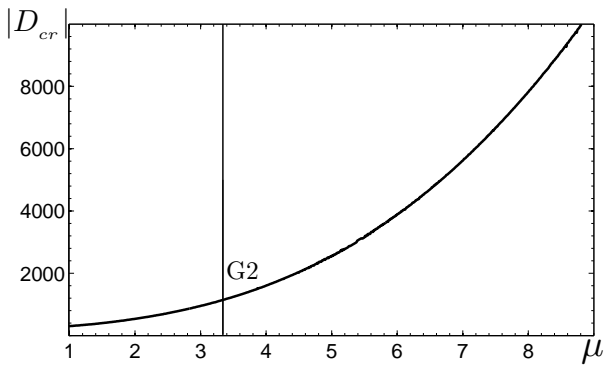}
\caption{\label{Fig3}
The dependence of the absolute value of the critical dynamo number, $|D_{\rm cr}|$ versus the parameter $\mu$. The vertical line indicates the spectral class $G2$ of the solar-like stars.}
\end{figure}

\begin{figure}
\centering
\includegraphics[width=9cm]{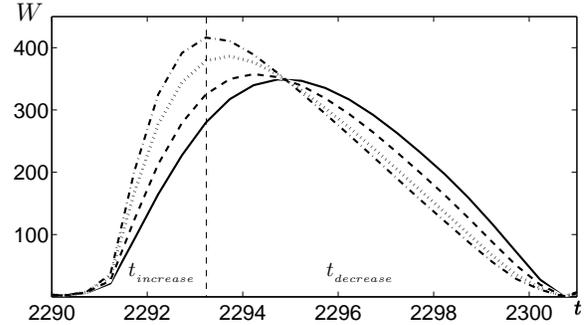}
\caption{\label{Fig4}
The time evolution of the Wolf number obtained from the numerical simulations for different values of the parameter $C_s$ entering in Eq.~(\ref{B4}): $C_s=0$ (solid), $C_s=1.4 \times 10^{-3}$ (dashed), $C_s=2.9 \times 10^{-3}$ (dotted) and $C_s=4.3 \times 10^{-3}$ (dashed-dotted).}
\end{figure}

We solve numerically Eqs.~(\ref{eqB}), (\ref{eqA}), (\ref{E9}) and~(\ref{B1}).
First, we study the properties of the dynamo model, i.e., we determine the parameter range for different regimes of the nonlinear oscillations of magnetic energy.

\subsection{Properties of the dynamo model}

In Fig.~\ref{Fig1} we show the dynamo number $-D$ versus the density parameter $\sigma$. We found three characteristic parameter ranges for different regimes of the nonlinear oscillations of magnetic energy, $E_B=\int B^2 \sin \theta \,d\theta \,d\phi$:

1. The oscillations with a constant frequency (Fig.~2a);

2. The oscillations with a low-frequency-modulation of amplitude and frequencies (Fig.~2b);

3. The chaotic behavior of magnetic energy (Fig.~2c).

It is clear that for small $\sigma$ the role of the dynamic nonlinearity is minor, and the observed nonlinear oscillations are regular with a constant frequency. The second regime of the nonlinear oscillations is associated with a transition between regular and chaotic behavior of the magnetic energy.
The third regime of the non-linear oscillations is the chaotic behavior of magnetic energy that may be a reason for complicated dynamics of the solar activity. In Fig.~\ref{Fig2} the time is normalized by 109 years to get the high-frequency oscillation period that is of the order of 11 years. Note that the radial turbulent diffusion time, $\mu^{-2} R_\odot^2 / \eta_{_{T}}$, is about $\sim 10$ years.
We remind also that the magnetic energy is averaged over a time that is of the order of 10 cycles of the dynamo waves (i.e., $\sim 109$ years).

For systems with a chaotic behaviour, the dynamics of different characteristics, like the magnetic energy, the poloidal and toroidal magnetic fields, and the magnetic helicity, is strongly dependent on initial conditions. To find a regime describing the observed dynamics of the parameters, one needs to perform a large number of numerical simulations. One of the most important observed parameters of solar activity measured during the last 270 years is the Wolf number \citep{G73,S89}. To determine the Wolf number in our model, we use Eq.~(\ref{B3}).

In Figs.~\ref{Fig3} and~\ref{Fig4} we show the properties of the dynamo model for different parameters. In particular, in Fig.~\ref{Fig3} we plot the dependence of the absolute value of the critical dynamo number, $|D_{\rm cr}|$ versus the parameter $\mu$. The vertical line indicates the spectral class $G2$ of solar-like stars. The value of the critical dynamo number $|D_{\rm cr}|>10^3$ is usually obtained for solar dynamo models and it weakly depends on the $\theta$-profiles. In Fig.~\ref{Fig4} we show
the time evolution of the Wolf number obtained through numerical simulations of our dynamo model for different values of the parameter $C_s$ entering in Eq.~(\ref{B4}) for the decay time of sunspots.

\subsection{Time evolution of Wolf number}

\begin{figure}
\centering
\includegraphics[width=9cm]{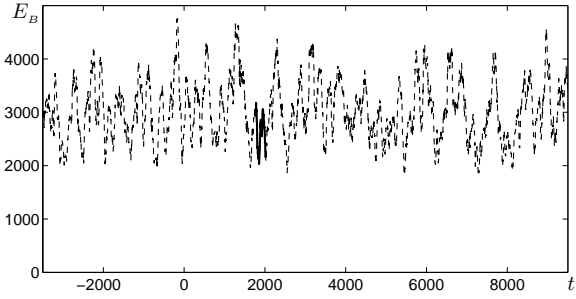}
\caption{\label{Fig5} The time evolution of the magnetic energy obtained in the numerical simulations of the dynamo model. The simulation time is $1.1 \times 10^4$ years. Black line is the part of the numerical simulation curve
that yields 70~\% correlation between observed and simulated Wolf numbers.}
\end{figure}

\begin{figure}
\centering
\includegraphics[width=9cm]{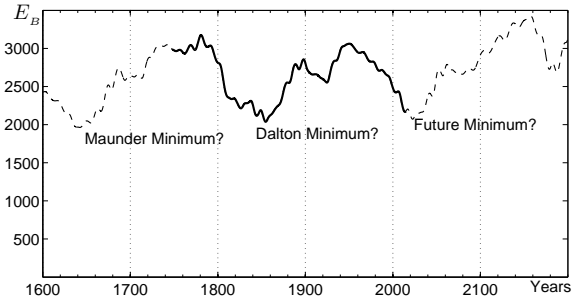}
\caption{\label{Fig6} The time evolution of the magnetic energy obtained in the numerical simulations of the dynamo model. The part of the numerical simulation curve of the time evolution of the magnetic energy that yields 70~\% correlation between observed and simulated Wolf numbers.}
\end{figure}

In Fig.~\ref{Fig5} we show the time evolution of the magnetic energy obtained through numerical simulations of the dynamo model over $1.1 \times 10^4$ years of simulation time. The parameters of the numerical simulation are as follows: $D=-8450$, $\sigma=3$, $\mu=3$, $\xi=0.1$, $\kappa_{_{T}}=0.1$, $R_\alpha=2$, $T=6.3$, $S_1=0.051$, $S_2=0.95$, where we use
the following initial conditions: $B_\phi(t=0,\theta)=S_1 \sin\theta + S_2 \sin(2\theta)$ and $A(t=0,\theta)$. The black line in Fig.~\ref{Fig5} is the part of the numerical simulation curve that yields a 70~\% correlation between observed and simulated Wolf numbers (see also the black line in Fig.~\ref{Fig6}). In Fig.~\ref{Fig6} we also indicate the Maunder and Dalton minima. This curve also indicates a possible minimum of solar activity in the near future.

\begin{figure}
\centering
\includegraphics[width=8cm]{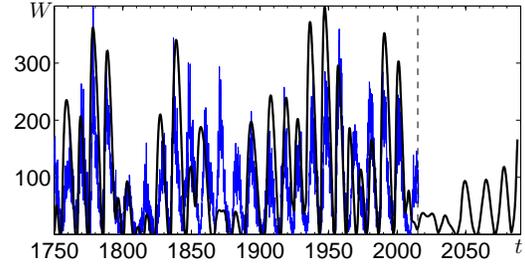}
\caption{\label{Fig7} The time evolution of the Wolf numbers obtained from numerical simulations (black curve) and observations (blue curve). The vertical dashed line indicates the year 2015.}
\end{figure}

A comparison between the time evolution of Wolf numbers obtained from the numerical simulations (black curve) and observations (blue curve) is shown in Fig.~\ref{Fig7}.
The correlation between numerical simulations and observations is about 70 \%. Note that the data for the Wolf number obtained from numerical simulations has not been averaged in time, contrary to the magnetic energy data.

\begin{figure}
\centering
\includegraphics[width=9cm]{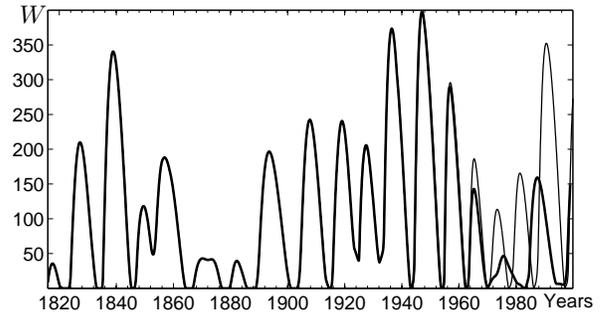}
\includegraphics[width=9cm]{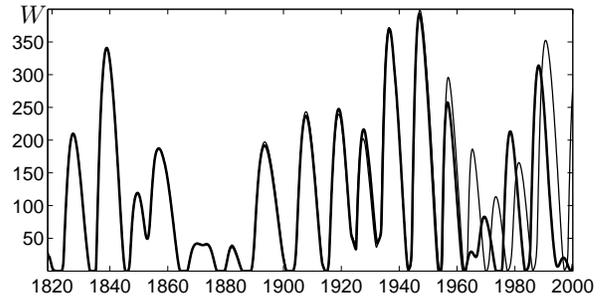}
\caption{\label{Fig8} The memory effect: the separation of two solutions of the dynamo system with very close initial conditions and the dependence of the memory time on the phase of the cycle, i.e., in what part of phase of solar activity (with increasing or decreasing solar activity) the solutions start their joint evolution.}
\end{figure}

An interesting question regards the memory time in the discussed dynamo system. By definition, the memory time describes the time interval in which two solutions with very close initial conditions are separated as they evolve. For example,
Figs.~\ref{Fig8} (upper and lower panels) show the time evolution of the Wolf numbers obtained from numerical solutions of the dynamo equations, for two cases with very close initial conditions. In Fig.~\ref{Fig8} (upper panel) two solutions start their joint evolution in the phase of increased solar activity and visible differences between the solutions are seen only after 145 years. In Fig.~\ref{Fig8} (lower panel) the two solutions start their joint evolution in the phase of decreased solar activity, and the small separation of these two solutions appears already after 70 years, while visible differences are seen only after 140 years. This implies that the memory time in the first case is 145 years (see Fig.~\ref{Fig8}, upper panel), and in the second case the memory time is 70 years (see Fig.~\ref{Fig8}, lower panel).

\begin{figure}
\centering
\includegraphics[width=9cm]{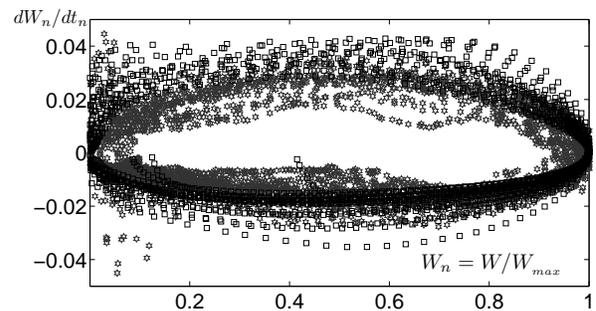}
\caption{\label{Fig9} The limit cycle for the derivative $dW_n/dt_n$ versus $W_n=W^{(n)}/W^{(n)}_{\rm max}$, where $n$ is the the number of the cycle obtained through numerical simulations (squares) and observations (stars).}
\end{figure}

\begin{figure}
\centering
\includegraphics[width=9cm]{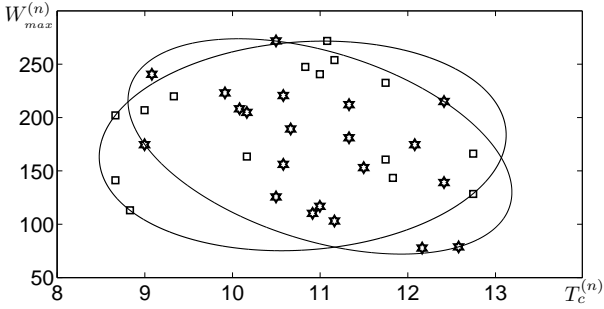}
\caption{\label{Fig10} The dependence of the maximum value of the Wolf number, $W^{(n)}_{\rm max}$ in the cycle $n$ versus the period, $T^{(n)}_c$ of this cycle obtained from numerical simulations (squares) and observations (stars). Ellipse shows the range of the random scattering of the function $W^{(n)}_{\rm max}(T^{(n)}_c)$.}
\end{figure}

For a more detailed study of the chaotic behaviour of the solar activity in
Fig.~\ref{Fig9} we plot the limit cycle in phase space, i.e., we show the function $dW_n/dt_n$ versus $W_n=W^{(n)}/W^{(n)}_{\rm max}$, where $W^{(n)}$ is the Wolf number in the cycle $n$ normalised by the maximum value of the Wolf number in this cycle, $W^{(n)}_{\rm max}$. The time $t_n$ is normalised here by the period, $T^{(n)}_c$ of this cycle. This figure demonstrates that in spite of strong variability caused by the chaotic behaviour, the solar cycle obeys a deep self-similarity. Similar behaviour can be observed in
Fig.~\ref{Fig10}, where we show the maximum value of the Wolf number versus the period, $T^{(n)}_c$, obtained through numerical simulations and observations. The ellipse indicates the range of random scattering of the function $W^{(n)}_{\rm max}(T^{(n)}_c)$. This figure also shows that the numerical modelling of Wolf numbers is in agreement with observations.

\begin{figure}
\centering
\includegraphics[width=8.5cm]{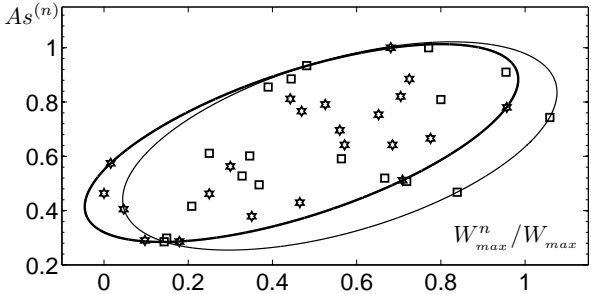}
\caption{\label{Fig11} Asymmetry ${\rm As}^{(n)}=t_{\rm decrease}/t_{\rm increase}$ obtained from numerical simulations (squares) and observations (stars). Here $W_{\rm max}$ is the maximum value of the Wolf number during the all observational time.}
\end{figure}

The asymmetry ${\rm As}^{(n)}=t_{\rm decr}/t_{\rm incr}$ of solar cycles obtained from numerical simulations and observations is plotted in Fig.~\ref{Fig11}, where $t_{\rm incr}$ (or $t_{\rm decr}$) is the instant in which the Wolf number increases (or decreases) with time. The form of the ellipse indicates that solar cycles with maximum solar activity (i.e., with maximum Wolf numbers) have the largest asymmetry. The overlap of the ellipses shown in Fig.~\ref{Fig11}, which correspond to the numerical modelling and observations, is about 90 \%.
Note that the dependence of the asymmetry of solar cycles on the amplitude of the cycle has been recently studied \citep{PK11,H15}.

\subsection{Variations of the parameters}

\begin{figure}
\centering
\includegraphics[width=8.5cm]{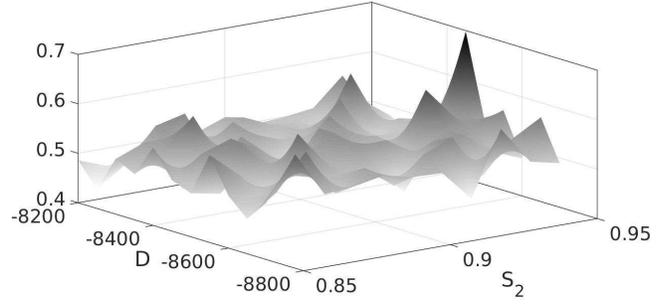}
\caption{\label{Fig12} The correlations between the numerical simulation data
for the Wolf number and the observational data depending on the variations of the parameters $D$ and $S_2$.}
\end{figure}

\begin{figure}
\centering
\includegraphics[width=8.5cm]{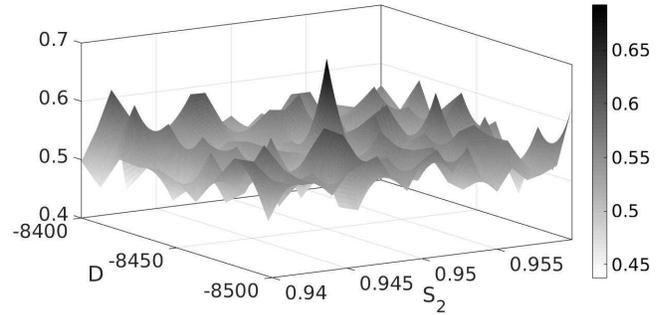}
\caption{\label{Fig13} The same as Fig.~\ref{Fig12} but for another range of the variations of the parameters $D$ and $S_2$.}
\end{figure}

In this section we discuss how the variations of the parameters affect the results.
There are two crucial parameters, the dynamo number $D$ and the initial field $B_{\rm init}^{\rm dip}$ for the dipole mode (determined by the parameter $S_2$), which strongly affect the dynamics of the nonlinear dynamo system.
The correct value of the initial field $B_{\rm init}^{\rm dip}$ allows us
to avoid very long transient regimes to reach the strange attractor.
In Figs.~\ref{Fig12} and~\ref{Fig13} we show the correlations between the numerical simulation data for the Wolf number and the observational data depending on the variations of the parameters $D$ and $S_2$. The maximum correlations is obtained when the parameters $D=-8450$
and $S_2=0.95$.

The dependence $D(\sigma)$ determines the region of the chaotic behaviour
(see region~3 in Fig.~\ref{Fig1}). As follows from Fig.~\ref{Fig1}, the value of the parameter $\sigma$ cannot be small, otherwise the dynamo system cannot remain inside region~3.
As follows from Fig.~\ref{Fig3} the parameter $\mu$
determines the critical dynamo number, $|D_{\rm cr}|$,
so that when $|D| > |D_{\rm cr}|$, the large-scale dynamo instability is excited.

The flux of the magnetic helicity, that is determined by
the parameter $\kappa_{_{T}}$, cannot be very small in order to avoid the catastrophic
quenching of the $\alpha$ effect. The optimal value
for this parameter is $\kappa_{_{T}} \approx 0.1$.
The variations of the other parameters only weakly affect
the obtained results.

\section{Conclusions}

To investigate the solar activity, we use a very simple one-dimensional nonlinear dynamo model of Parker's dynamo waves. Magnetic fields depend on time and co-latitude. We take into account the algebraic nonlinearity of the alpha-effect and the dynamic nonlinearity caused by the evolution of the magnetic helicity. The dynamic nonlinearity causes complicated behavior of solar activity for large dynamo numbers. To simplify the model we account for the turbulent diffusion flux of the magnetic helicity. To describe the solar activity we use a phenomenological equation for the evolution of Wolf numbers. This equation takes into account the mechanism of formation of sunspots, based on the negative effective magnetic pressure instability (NEMPI), and the decay of sunspots in a nonlinear evolution of NEMPI. In particular, to determine the rate of production of sunspots we take into account the growth rate of NEMPI and the fact that this mechanism does not create new magnetic flux, but rather it redistributes the magnetic flux produced by the mean-field dynamo. This mechanism creates magnetic spots (strong magnetic concentrations) in small areas of solar surface.

We use the no-z dynamo model, rather than a more sophisticated three-dimensional dynamo model, since the information about spatial distribution of the kinetic alpha effect is not known from observations.
There is also strong dependence of the rotation (characterized by the Coriolis number) on the depth of the convective zone (from very slow rotation at the surface of the sun to fast rotation near the bottom of the convective zone. Since the plasma density varies over seven orders of magnitude from the top to the bottom of the convective zone, it introduces additional complication in the modelling of the solar activity. This implies that an application of a more sophisticated three-dimensional dynamo model with unknown from observations coefficients for the prediction of the solar activity seems to be not improve a quality of the prediction.

Numerical simulations of our dynamo model with a phenomenological equation for the evolution of the Wolf number demonstrates good agreement between numerical modelling and observations (with about a 70 \% correlation in observed data and simulations of Wolf numbers). In particular, we determine the dependence of the maximum value of the Wolf number in the given cycle versus the period of this cycle. We also find the asymmetry of the solar cycles versus the amplitude of the cycle. The numerical modelling performed in this study cannot directly serve as an instrument for prediction of solar activity because we have not used any procedure similar to the real date assimilation as used, for instance, in the atmospheric weather prediction. However, this model describes the general features of solar activity. In particular, it indicates the decreased level of solar activity in the near future (see Figs.~\ref{Fig6}-\ref{Fig7}).

\section*{Acknowledgments}

Financial support from the Research Council of Norway
under the FRINATEK grant 231444 (NK, IR),
from grant "Methods of analysis and data processing,
visualisation and prediction
of multi-dimensional data" (NS, SP),
from the RFBR support under grant 15-02-01407 (DS)
are gratefully acknowledged.
The authors (NK, IR, DS) thank NORDITA for hospitality during their visits.
NK thanks Ural Federal University for hospitality during his visits.

\end{document}